\documentclass[aps,pra,showpacs,twocolumn,amsmath,amssymb,floatfix]{revtex4}
\pdfoutput=1

\usepackage{graphicx,subfigure}
\usepackage{bm}
\usepackage{xspace}
\usepackage[latin1]{inputenc}

\graphicspath{{../figures/}{../figures/converted/}}

{\catcode`\|=\active 
\gdef\Braket#1{\left<\mathcode`\|"8000\let|\BraVert {#1}\right>}}
\def\BraVert{\egroup\,\vrule\,\bgroup}
\newcommand{\bra}[1]{\left\langle{#1}\right\rvert}
\newcommand{\ket}[1]{\left\lvert{#1}\right\rangle}

\renewcommand{\vec}[1]{\ensuremath{\boldsymbol{#1}}}
\newcommand{\vhat}[1]{\ensuremath{\hat{\vec{#1}}}}

\DeclareMathOperator*{\re}{Re} 
\DeclareMathOperator*{\im}{Im}

\newcommand{\topp}[1]{^{({#1})}}

\newcommand{\belowline}[2]{\begin{minipage}[t]{#1}\vspace{0pt}\par{#2}\end{minipage}}
\newcommand{\belowlabel}[1]{\raisebox{-1ex}[0pt][0pt]{\makebox[0pt][l]{{#1}}}}

\newcommand{\ree}{\rho_\text{ee}}
\newcommand{\deff}{\delta_\text{eff}}

\newcommand{\erel}{\varepsilon}
\newcommand{\rrel}{r}
\newcommand{\trel}{\tau}
\newcommand{\reerel}{R_\mu}
\newcommand{\eofmaxdndt}{\erel_s}
\newcommand{\eofmaxdedt}{\erel_c}

\newcommand{\drel}{\delta}
\newcommand{\dreld}{\delta_D}
\newcommand{\dreldi}[1]{\delta_D\topp{#1}}
\newcommand{\dmax}{\delta_M}
\newcommand{\dmaxi}[1]{\delta_M\topp{#1}}

\newcommand{\eprop}{\Xi}
\newcommand{\pho}{P_D}

\newcommand{\tint}{T_\text{tot}}
\newcommand{\amm}{a}
\newcommand{\rbar}{\bar{R}}

\newcommand{\pdz}{P_D\topp{z}}

\newcommand{\ke}{\ket{\text{e}}}
\newcommand{\kg}{\ket{\text{g}}}
\newcommand{\be}{\bra{\text{e}}}
\newcommand{\bg}{\bra{\text{g}}}
\newcommand{\wlaser}{\nu_\text{Laser}} 
\newcommand{\wres}{\nu_{ge}}

\newcommand{\xavg}{\bar{\vec{x}}}
\newcommand{\xiavg}{\bar{x}_i}

\begin{document}

\title{Fluorescence during Doppler cooling of a single trapped atom}

\author{J. H. Wesenberg}
\email{janus.wesenberg@nist.gov}
\author{R. J. Epstein}
\author{D. Leibfried}
\author{R. B. Blakestad}
\author{J. Britton}
\author{J. P. Home}
\author{W. M. Itano}
\author{J. D. Jost}
\author{E. Knill}
\author{C. Langer}
\altaffiliation[Present address: ]{Lockheed Martin, Huntsville, AL}
\author{R. Ozeri}
\altaffiliation[Present address: ]{Weizmann Institute of Science, Rehovot, Israel}
\author{S. Seidelin}
\author{D. J. Wineland}

\affiliation{National Institute of Standards and Technology, Boulder, Colorado 80305}

\date{\today}

\begin{abstract}
  We investigate the temporal dynamics of Doppler cooling of an
  initially hot single trapped atom in the weak binding regime using a
  semiclassical approach.
  We develop an analytical model for the simplest case of a single
  vibrational mode for a harmonic trap, and show how this model allows
  us to estimate the initial energy of the trapped particle by
  observing the fluorescence rate during the cooling process.
  The experimental implementation of this temperature measurement
  provides a way to measure atom heating rates by observing the
  temperature rise in the absence of cooling.  This method is
  technically relatively simple compared to conventional sideband
  detection methods, and the two methods are in reasonable agreement.
  We also discuss the effects of RF micromotion, relevant for a
  trapped atomic ion, and the effect of coupling between the
  vibrational modes on the cooling dynamics.
\end{abstract}


\maketitle

Laser cooling of trapped neutral atoms and atomic ions is a well
established technique: for example, cooling to the motional ground
state
\cite{diedrich89:laser_coolin_to_zero_point,monroe95:resol_sideb_raman_coolin_bound,perrin98:sideb_coolin_neutr_atoms_in}
and motional state tomography \cite{leibfried98:shadow_and_mirror} are
routinely performed with resolved motional-sideband excitation
techniques.
Sideband techniques require the natural linewidth $\Gamma$  of the cooling
transition to be small compared to the vibrational frequency of the
trapped particle, in order to allow the motional sidebands to be resolved.
%
Many experiments are, however, conducted in the ``weak-binding
regime'', where $\Gamma$ is larger than the oscillation
frequency. Here, the cooling process is essentially the same as
Doppler cooling of free atoms, because the spontaneous decay process
is short compared to the atom's oscillation period
\cite{wineland79:laser_coolin_atoms}.
Even in experimental setups that implement sideband techniques, an
initial stage of such ``Doppler cooling'' is often employed.
The first examinations of Doppler cooling of trapped ions \cite{%
wineland78:radiat_press_coolin_bound_reson,%
neuhauser78:optic_sideb_coolin_visib_atom,%
wineland79:laser_coolin_atoms,%
javanainen80:laser_coolin_trapp_partic_i,%
javanainen81:laser_coolin_trapp_partic_ii,%
javanainen81:laser_coolin_trapp_partic_iii,%
itano82:laser_coolin_ions_stored_in%
} did not take
into account the effects of micromotion due to the trapping RF field.
After cooling and heating effects related to micromotion were observed,
these effects were explained theoretically
\cite{bluemel89:chaos_and_order_laser_cooled,devoe89:role_laser_dampin_trapp_cryst,cirac94:laser_coolin_trapp_ions}
by including the effects of micromotion.

Here, we consider Doppler cooling of a single trapped atom or ion.
While most previous work has focused on the final stages of cooling,
our focus will be on the temporal dynamics of the cooling,
particularly in the ``hot regime'' where the Doppler shift due to atom
motion is comparable to or much larger than $\Gamma$.
For the 1-D case we find that the cooling rate can be calculated
analytically in the weak-binding regime without assuming the atom to
be in the Lamb-Dicke regime.
For a trapped ion, when we take RF micromotion into consideration,
stable, highly excited states emerge when only one mode is considered
\cite{peik99:sideb_coolin_ions_in_radio}. When all three vibrational
modes of the ion are considered, we find that couplings between the
modes tend to break the stability of such points allowing cooling to
reach the Doppler limit.

A practical application of our results is to estimate the initial
motional energy of an atom or ion from observations of the
time dependence of the fluorescence during the cooling process.
As mentioned above, sideband spectroscopy is the conventional
technique for characterizing motional states, and it has been used to
characterize the heating rate of ions in the absence of cooling
\cite{diedrich89:laser_coolin_to_zero_point,monroe95:resol_sideb_raman_coolin_bound,turchette00:heatin_trapp_ions_from_quant,seidelin06:microf_surfac_elect_trap_scalab,deslauriers06:scalin_suppr_anomal_quant_decoh,pearson06:exper_inves_planar_ion_traps,epstein07:simpl_ion_heatin_rate_measur}.
However, it is more complicated to implement experimentally than
Doppler cooling, requiring more laser beams.
%
Currently, considerable effort is being devoted to understanding the
anomalous heating observed in ion traps
\cite{turchette00:heatin_trapp_ions_from_quant,seidelin06:microf_surfac_elect_trap_scalab,deslauriers06:scalin_suppr_anomal_quant_decoh,pearson06:exper_inves_planar_ion_traps,epstein07:simpl_ion_heatin_rate_measur}.
A less complicated technique for measuring temperature could simplify
this work.

This paper is structured as follows: 
In Sec.~\ref{sec:model} we present a semiclassical model of the Doppler
cooling process for a bound atom in the weak-binding regime.  
In Secs.~\ref{sec:analyzis} and \ref{sec:thermal-averaging} we analyze
the fluorescence predicted by the model for a single vibrational mode
unaffected by micromotion. Here, we consider single cooling
trajectories and average over these with a given distribution of
initial motional energies.
We derive expressions useful for estimating initial temperature from
fluorescence observations in these sections.
Sec.~\ref{sec:optim-exper-param} discusses how to minimize the total
measurement time required to estimate the mean initial energy.
In Sec.~\ref{sec:effects-spect-modes} we consider the effects of other
motional modes with and without taking into account any RF micromotion
experienced by such modes.
Sec.~\ref{sec:modif-exper-prot} suggests modifications to the basic
experimental protocol that might provide improved sensitivity of the
temperature measurements.
Sec.~\ref{sec:conclusion} concludes the paper.

\section{Model}
\label{sec:model}

We consider a semiclassical model of Doppler cooling of a single
weakly trapped atom
\cite{wineland79:laser_coolin_atoms,itano82:laser_coolin_ions_stored_in}.
We will initially consider only a single mode of motion, taken to be
along the $z$ direction.  We assume a harmonic potential with
oscillation frequency $\omega_z$.
In Sec.~\ref{sec:effects-spect-modes} we consider a more detailed
model that includes three dimensions and micromotion for ions.

The atom is Doppler-cooled by a single laser beam of angular
frequency $\wlaser$ and wave-vector $\vec{k}$, detuned by
$\Delta\equiv \wlaser-\wres$ from the resonance frequency $\wres$ of a
two-level, or ``cycling'', transition between two internal states,
$\kg$ and $\ke$, of the atom. We write the coupling Hamiltonian as
\begin{equation}
  \label{eq:couplingh}
  H\topp{c}=
  \hbar\, \Omega_\text{Rabi} 
  \,\left(\ke\bg+\kg\be\right)
  \,\cos(\vec{k}\cdot\vec{x}-\wlaser t),
\end{equation}
where $\vec{x}$ is the atom position, $2\pi\hbar$ is Planck's
constant, and $\Omega_\text{Rabi}$ is the resonant Rabi frequency.

We assume the atom is weakly bound in the $z$ direction, that is,
$\omega_z$ is much less than the excited state decay rate $\Gamma$.
The atom's level populations are then approximately in steady
state with respect to the instantaneous effective detuning,
$\Delta_\text{eff}\equiv\Delta+\Delta_D$, including the Doppler shift
$\Delta_D\equiv-k_z v_z$, where $v_z$ and $k_z$ are the $z$-components of the
velocity and wave-vector. The excited state population is then \cite{Loudon73}
\begin{equation}
  \ree(v_z)=\frac{s/2}{1+s+(2 \Delta_\text{eff}/\Gamma)^2}.
\end{equation}
Here $s$ is the saturation parameter, proportional to the cooling beam
intensity, $s\equiv2 |\Omega_\text{Rabi}|^2/\Gamma^2$.

The excited state population is associated with the photon scattering 
rate $dN/dt$ by the relation $dN/dt=\Gamma \ree(v_z)$. 
While the momentum kicks associated with photon emission are assumed
to average to zero over many absorption-emission cycles, the absorbed
photons will impart a velocity dependent momentum transfer due to the
scattering that can be described by a velocity-dependent force
\begin{equation}
  \label{eq:fracd-vecpd-t}
  F_z(v_z)=m \frac{d v_z}{d t}= \hbar k_z \, \Gamma \ree(v_z),
\end{equation}
where $m$ is the atom's mass.
This velocity-dependent force will in general change the motional
energy $E$ of the atom. If the relative change in energy over a motional
cycle is small, we can average the effect of $F_z$ over the
oscillatory motion to find the evolution of $E$:
\begin{equation}
  \label{eq:dedtth}
  \frac{dE}{dt}= \left\langle v_z F_z(v_z) \right\rangle,
\end{equation}
where the average is over one motional cycle.
The average energy change per scattering event is $dE/dN = \hbar k_z
v_z = -\hbar \Delta_D$.

In addition to $F_z$, the atom will experience a stochastic force due
to photon recoil that, assuming isotropic emission, will cause
heating at a rate \cite{itano82:laser_coolin_ions_stored_in}
\begin{equation}
  \label{eq:recoil}
  \left(\frac{dE}{dt}\right)_\text{recoil}=\frac{4}{3} 
  \frac{\left(\hbar\, k_z\right)^2}{2 m} 
  \, \frac{dN}{dt},
\end{equation}
where $(\hbar k_z)^2/2 m$ is the recoil energy associated with the
scattering. We will mostly ignore the effects of recoil heating in
what follows since it will be important only near the cooling limit.

\section{Analysis}
\label{sec:analyzis}
We will now analyze the time-dependence of the atom fluorescence 
during the Doppler cooling process, as predicted by the model
introduced above.

To simplify the algebra, we will scale energies by $\hbar$ times half
the power-broadened linewidth, and time by the resonant scattering
rate:
\begin{subequations}
  \label{eq:units}
\begin{align*}
  \{ \erel ,\drel, \rrel \} 
  &= \{E,\hbar \Delta, \frac{(\hbar  k_z)^2}{2 m} \} / E_0, 
  &E_0&=\frac{\hbar \Gamma}{2} \sqrt{1+s}\\
  \trel                     
  &= t/t_0,                                                
  &t_0&=\left(\Gamma \frac{s/2}{1+s}\right)^{-1}.
\end{align*}
\end{subequations}
%
%
%
%
As an example of typical values, we consider a trapped
$^{25}\text{Mg}^+$ ion, where the $^2S_{1/2} - ^2P_{3/2}$ cooling
transition at $279.6\,\text{nm}$ has a natural linewidth of $\Gamma =
2 \pi\times 41.4\,\text{MHz}$. At a detuning of $\Delta=-2 \pi\times
20 \,\text{MHz}$ with $s=0.9$ and $k_z/k=0.71$, we find that
$E_0/k_B=1.4\,\text{mK}$, where $k_B$ is Boltzmann's constant, and
$t_0=16\, \text{ns}$.  The detuning and recoil parameters are
$\drel=-0.70$ and $\rrel=0.0018$.

\begin{figure}
  \centering
  \includegraphics[width=0.95\linewidth]{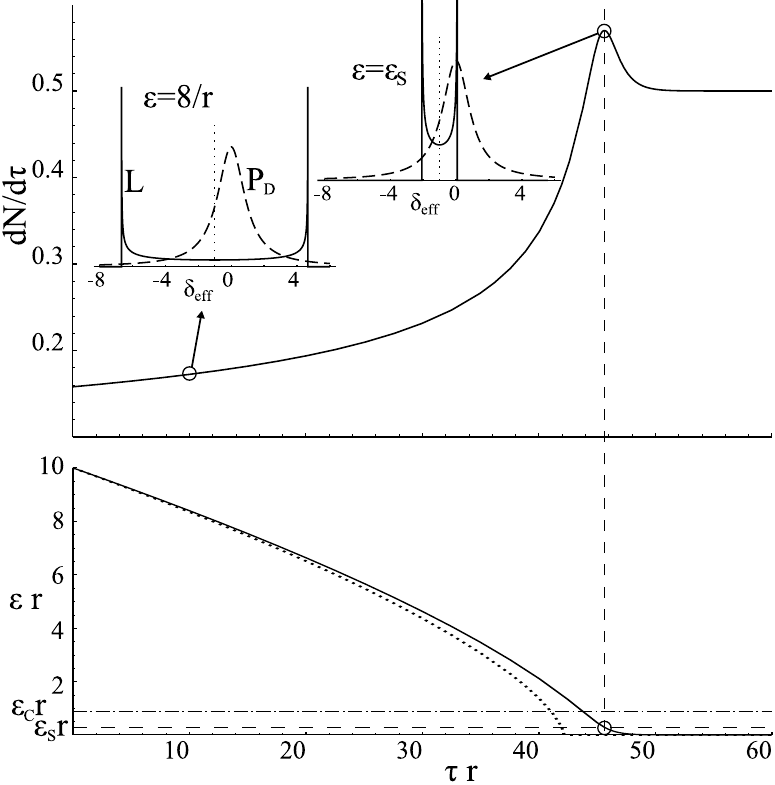}
  \caption{ %
    Scattering rate (top) and energy (bottom) as a function of time
    during the Doppler cooling of a single atom from an initial energy
    of $10/r$, as given by Eqs.~\eqref{eq:dedtres} and
    \eqref{eq:dndtrel} at a detuning of half the power-broadened
    linewidth ($\delta=-1$).
    The dotted curve in the lower plot shows the energy predicted by
    the asymptotic approximation \eqref{eq:ereloftapr}.
    Insets in the upper plot show (at two different times) the two components appearing in the
    integral defining $dN/d\trel$ in Eq.~\eqref{eq:dndtrel}: The
    probability density of the effective detuning
    $\pho(\dmax;\deff-\drel)$ (solid) and the Lorentzian line
    profile, $L(\deff)=1/(1+\deff^2)$ (dashed) as a function of $\deff=\drel+\dreld$ at
    $\erel =8/\rrel$ and $\erel=\eofmaxdndt$.
    Scattering events where the atom is moving towards the laser so
    that $\dreld>0$, corresponding to the rightmost
    peak of $\pho$, result in cooling, and vice versa.
    The energies of maximal cooling and scattering rates,
    $\eofmaxdedt$ and $\eofmaxdndt$, are given by
    Eqs.~\eqref{eq:esteep} and \eqref{eq:ebright}.  
  }
  \label{fig:singleion}
\end{figure}

The maximal change in energy per scattering event at a given energy is
$\dmax\equiv 2 \sqrt{\erel \rrel}$.
The energy at which the maximal Doppler shift, which in the scaled
units is equal to $\dmax$, is equal to the power-broadened linewidth,
$\dmax=2$, is of interest during the cooling process.
%
%
For reference we note that this energy corresponds to
\begin{equation}
  \label{eq:escale}
  \frac{1}{r}\,E_0 = (1+s) \frac{\hbar^2 \Gamma^2}{4} \frac{2
    m}{\hbar^2 k_z^2}.
\end{equation}
For the typical experimental parameters considered above, 
$E_0/r$ is equal to $k_B \times 700\,\text{mK}$ or $3700\, \hbar
\omega_z$ for $\omega_z=2\pi\times 4.0\, \text{MHz}$.

For harmonic oscillations, the
instantaneous Doppler shift $\dreld\equiv \hbar\Delta_D/E_0$ is
distributed according to the probability density
\begin{align}
  \label{eq:pho}
  \pho(\dmax;\dreld)
  &=\int_0^{2\pi} \delta_\text{Dirac}(\dreld-\dmax \sin(\phi))\, \frac{d\phi}{2\pi}\notag\\
  &=
  \begin{cases}
    \frac{1}{\pi} \frac{1}{\sqrt{\dmax^2-\dreld^2}}  &\text{if } \lvert
    \dreld\rvert<\dmax,\\
    0&\text{otherwise,}    
  \end{cases}
\end{align}
where $\delta_\text{Dirac}$ is the Dirac $\delta$ function.
Since the average energy change per scattering event is $-\dreld$, and
the instantaneous scattering rate is $1/(1+\deff^2)$, where
$\deff\equiv\drel+\dreld$, the rate of change of $\erel$ averaged over
the secular oscillations by Eq.~\eqref{eq:dedtth} takes the form:
\begin{equation}
  \label{eq:dedtrel}
  \frac{d\erel}{d\trel}=\int 
  -\dreld   
  \pho(\dmax;\dreld) 
  \frac{1}{1+(\drel+\dreld)^2}
  d\dreld,
\end{equation}
as illustrated in Fig.~\ref{fig:singleion}.
We can evaluate the integral as detailed in Appendix
\ref{sec:integrals}, to find that
\begin{subequations}\label{eq:dedtrelboth}
\begin{align}
  \label{eq:dedtres}
  \frac{d\erel}{d\trel}&= \frac{1}{2 \sqrt{\erel \rrel}} (\re(Z)+\drel \im(Z))\\
  \label{eq:dedtresapr}
    &\approx \frac{\drel}{2 \sqrt{\erel \rrel}},\quad 
    \erel \gg  (1+\drel^2)/r,
\end{align}
\end{subequations}
where $Z=Z(\drel,\dmax)= i /\sqrt{1- (\drel+i)^2/4 \erel \rrel}$.
The asymptotic approximation \eqref{eq:dedtresapr} corresponds to
approximating $\pho(\dmax;\dreld)$ by $\pho(\dmax;0)$, which is
reasonable in the ``hot'' regime, where the peaks of $\pho$ have small
overlap with the Lorentz line profile.

The scattering rate averaged over the motion is analogous to
Eq.~\eqref{eq:dedtrelboth} and is given by
\begin{align}
  \label{eq:dndtrel}
  \frac{dN}{d\trel}&=\int \pho(\dmax;\dreld) \frac{1}{1+(\drel+\dreld)^2}
  d\dreld \notag\\
  &=\frac{1}{2 \sqrt{\erel \rrel}} \im(Z),
\end{align}
as illustrated in Fig.~\ref{fig:singleion}.  In the limit of
$\erel\gg(1+\drel^2)/r$, we find $dN/d\trel \approx 1/(2\sqrt{\erel
  \rrel})$, so that according to Eq.~\eqref{eq:dedtresapr} we have in
this limit $d\erel/dN \approx \drel$. This corresponds to each photon
on average extracting an energy of $\hbar \Delta$.
This can be understood by noting that in the limit of
$\erel\gg(1+\drel^2)/r$, $\pho(\dmax;\dreld)$ is to a good approximation
uniform over the Lorentz line profile, and so the value of
$\deff=\drel+\dreld$ averaged over the scattering events will be zero.
Since each scattering event extracts an energy of $-\dreld$, the
average cooling per scattering event should indeed be $\drel$.

\begin{figure}
  \centering
  \includegraphics[width=0.95\linewidth]{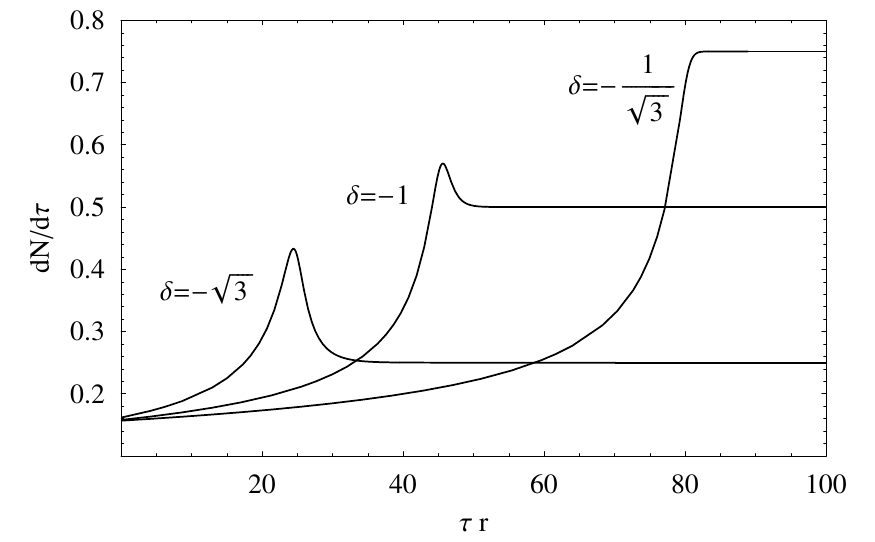}
  \caption{ 
    The scattering rate, $dN/d\trel$, vs.~time during Doppler cooling
    of a single atom with initial energy $\erel = 10/\rrel$ for
    different laser detunings.
    For $\drel<\drel_C=-1/\sqrt{3}$, a maximal scattering rate occurs
    at $\erel=\eofmaxdndt$, as given by Eq.~\eqref{eq:ebright}. The maximal
    value of the scattering rate is given by Eq.~\eqref{eq:srfactor}.
    Closer to resonance, the scattering rate increases monotonically 
    during the cooling.  
  }
  \label{fig:detuning}
\end{figure}

The time-dependence of $\erel$ is formally found by integrating
$d\erel/d\trel$ as given by Eq.~\eqref{eq:dedtrelboth}. For the
asymptotic approximation \eqref{eq:dedtresapr} we find
\begin{equation}
  \label{eq:ereloftapr}
  \erel(\trel) \approx \left(\erel_0^{3/2}+\frac{3 \drel \trel}{4
      \sqrt{\rrel}}\right)^{2/3}, \quad  \erel \gg  (1+\drel^2)/r,
\end{equation}
where $\erel_0$ is the energy at $\trel=0$, as plotted in the lower
part of Fig.~\ref{fig:singleion}.
For the exact expression \eqref{eq:dedtres}, we must resort to
numerical methods, although we do find analytically that the cooling
rate is maximal for $\erel$ related to $\drel$ by
\begin{equation}
  \label{eq:esteep}
  \eofmaxdedt=\frac{1+\drel^2}{2 r} \cos(\frac{1}{3} \arccos(\frac{1-\drel^2}{1+\drel^2})),
\end{equation}
which quantifies our previous observation that $1/r$ is a typical
energy scale of the cooling process.

The behavior of $dN/d\trel$ is qualitatively different for $\drel$
being smaller or larger than a critical detuning,
$\drel_C\equiv -1/\sqrt{3}$.  For $\drel < \drel_C$, $dN/d\trel$ has a
maximum at
\begin{equation}
  \label{eq:ebright}
  \eofmaxdndt=\frac{1}{4 r} (\drel-\sqrt{3})(\drel+1/\sqrt{3}).
\end{equation}
For the example parameters listed below Eq.~\eqref{eq:units},
$\drel_C$ corresponds to a detuning of $\drel_C E_0/\hbar=-2\pi\times16.5\,\text{MHz}$.
Closer to resonance, i.e., when $\drel_C<\drel<0$, no maximum occurs,
as illustrated in Fig.~\ref{fig:detuning}.
The maximal scattering rate is reached when one of the peaks of the
Doppler distribution \eqref{eq:pho} is in resonance with the
cooling transition, as illustrated in the insets of
Fig.~\ref{fig:singleion}.
In the regime where a maximum exists, the maximal scattering rate is
found to exceed the steady state scattering rate by a factor of 
\begin{equation}
  \label{eq:srfactor}
  \left.\frac{dN}{d\trel}\right\lvert_{\erel=\eofmaxdndt} 
  / 
  \left.\frac{dN}{d\trel}\right\lvert_{\erel=0} 
  = \frac{\sqrt{3\sqrt{3}}}{4} \frac{1+\drel^2}{\sqrt{|\drel|}}.  
\end{equation}

We emphasize that the only approximations made above are the
weak-binding approximation and the omission of recoil heating.  In
particular, the trapped particle is not assumed to be in the
Lamb-Dicke regime. For the weak-binding regime, $\dmax>1$
implies that the motion is well outside the Lamb-Dicke regime.
To find the cooling rate predicted by \eqref{eq:dedtrelboth} in the
Lamb-Dicke limit, we note that to first order in $\erel$ we have
$d\erel/d\trel \approx 4 \drel \erel \rrel/ (1+\drel^2)^2$.
This corresponds to $\erel$ decreasing exponentially with
$\trel$. Except for the omission of recoil heating, the value of the
decay time agrees with previous work that assumed the atom
was in the Lamb-Dicke regime
\cite{wineland79:laser_coolin_atoms,wells90:simpl_theor_sideb_coolin}.

In the above analysis, we have ignored recoil heating as given by
Eq.~\eqref{eq:recoil}.
In the limit of $\erel \gg \eofmaxdedt$, the ratio of heating to cooling is
seen to be $4 r/(3 |\drel|)$, which is a small fraction for realistic
parameters.
For $\erel < \eofmaxdedt$, the cooling is less efficient and the
contribution from recoil becomes more significant, leading to a
nonzero steady-state energy. Nevertheless, ignoring recoil heating is
reasonable when considering only fluorescence, since the scattering
rate has almost reached its steady-state value when the effect of
recoil becomes important.
We have omitted recoil in this analysis to make $\drel$ the only free
parameter and simplify the discussion. Recoil can be included in
calculations by combining Eqs.~\eqref{eq:recoil}, \eqref{eq:dedtres},
and \eqref{eq:dndtrel}.

\section{Thermal averaging}
\label{sec:thermal-averaging}

An application of the analysis presented above is to estimate the
initial motional energy of a trapped atom from the fluorescence
observed during the cooling process.
Using this method, we can estimate the average rate of heating
experienced by a trapped atom in the absence of cooling by first
allowing the atom to heat up without cooling for a certain period and
then observing the time dependence of the fluorescence as the atom is
re-cooled.
As discussed in Sec.~\ref{sec:analyzis}, we have for
$\erel\gg(1+\drel^2)/r$ that the average cooling per scattering event
is $\drel$. The approximate total number of photons scattered during
the cooling of an atom with initial motional energy
$\erel_\text{initial}$ can consequently be approximated by
$\lvert\erel_{\text{initial}}/\drel\rvert$.
For the example parameters given in Sec.~\ref{sec:analyzis}, this
corresponds to $\approx 800$ photons for
$\erel_\text{initial}=1/\rrel$, corresponding to $k_B\times
700\,\text{mK}$.  With typical photon detection efficiencies of less
than $10^{-3}$, very few photons are registered in a single
experiment.  We must therefore repeat many experimental cycles
consisting of a heating period and a cooling period.

We now consider the form of the fluorescence signal when averaged
over many such experimental cycles.
Here, we will assume the heating is stochastic and take the
distribution $P_0(\erel)$ of the motional energies at the beginning of
each cooling period to be the Maxwell-Boltzmann distribution with mean
energy $\bar{\erel}$,
\begin{equation}
  \label{eq:thermaldist}
  P_0(\erel)=\tfrac{1}{\bar{\erel}}\, e^{-\erel/\bar{\erel}}.
\end{equation}
However, the results below hold for any form of $P_0(\erel)$.

\begin{figure}
  \centering
  \includegraphics[width=\linewidth]{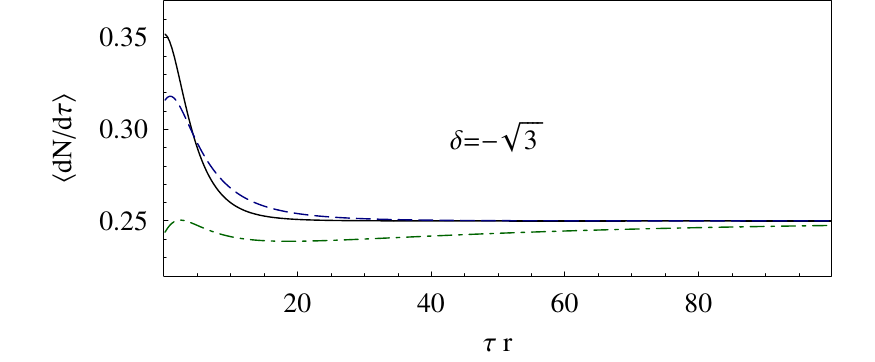}\\
  \includegraphics[width=\linewidth]{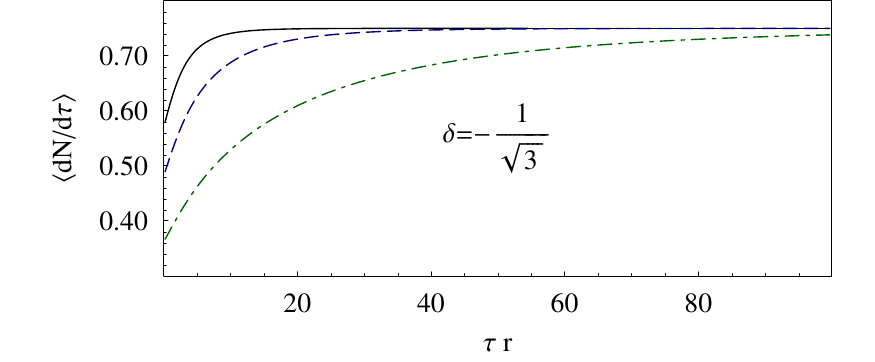}
  \caption{
    Thermally averaged scattering rate vs.~time, for $\drel=-\sqrt{3}$ (top) and
    $\drel=-1/\sqrt{3}=\drel_C$ (bottom).  In both figures, $P_0$ is assumed
    to be a thermal distribution with $\bar{\erel}$ equal to $1$,$2$,
    and $4$ times $\eofmaxdedt$ for the solid, dashed, and dash-dotted
    lines respectively.
    Note that for $\drel<\drel_C$, the initial fluorescence is larger
    than the steady-state fluorescence for low values of
    $\bar{\erel}$; this is attributed to the local maximum in the
    fluorescence illustrated in Fig.~\ref{fig:detuning}.
  }
  \label{fig:thermalavg}
\end{figure}

The thermally averaged scattering rate is conveniently written in
terms of the propagator, $\eprop$, of $\erel$: Let
$\eprop(\erel_0,\trel)$ denote the energy at time $\trel$ of an atom
with initial energy $\erel(\trel=0)=\erel_0$. We can then write the
thermally averaged scattering rate at time $\trel$ as
\begin{equation}
  \label{eq:dndtthermalnum}
  \left\langle \frac{dN}{d\trel} \right\rangle_{\bar{\erel}} 
  = \int_0^\infty P_0(\erel')  
  \left.\frac{dN}{d\trel}\right|_{\erel= \eprop (\erel',\trel)} 
  d\erel'.
\end{equation}
This can be efficiently computed numerically by noting that
$\eprop(\eprop(\erel,\trel_1),\trel_2)=\eprop(\erel,\trel_1+\trel_2)$,
as detailed in Appendix
\ref{sec:numer-eval-therm}. Figure \ref{fig:thermalavg} shows the
thermally averaged scattering rate for a few different parameters.

\begin{figure}
  \centering
  \includegraphics[width=8cm]{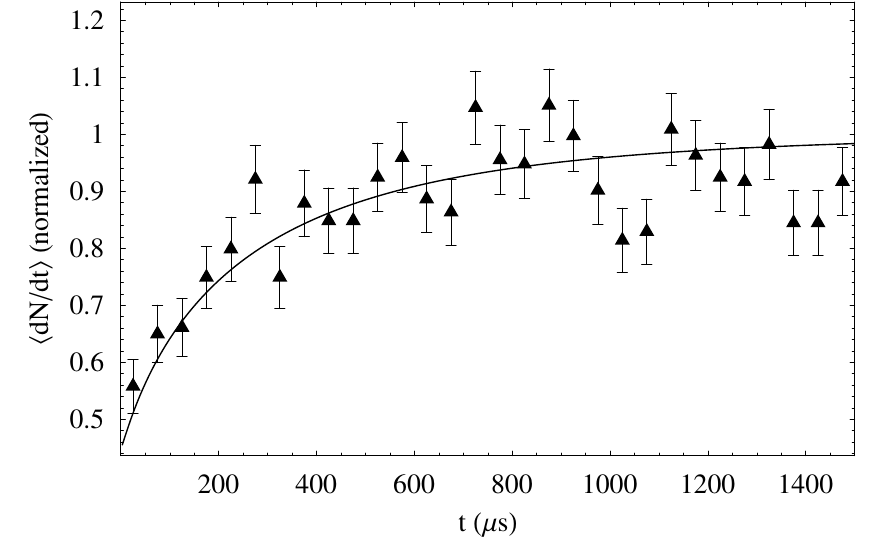}
  \caption{ Experimentally observed fluorescence during Doppler
    cooling of a single $\text{Mg}^+$ ion compared with the fluorescence
    predicted by the simple 1-D model.
    Data points (triangles) indicate the observed scattering rates, 
    obtained by integrating over many experiments.  In
    each experiment, the time-resolved fluorescence is recorded while
    the ion is re-cooling after having been allowed to heat up for a
    period of $25\,\text{s}$. The experimental parameters were those
    given after Eq.~\eqref{eq:units}.
    Error bars are based on counting statistics.  
    The solid curve is the scattering rate predicted by
    Eq.~\eqref{eq:dndtthermalnum}, assuming the motional energy of
    the ion after the heating period to be given the Maxwell-Boltzmann
    distribution \eqref{eq:thermaldist} with $\bar{\erel}=5.1/r$,
    corresponding to a temperature of $3.9\,\text{K}$.
    Since $\bar{\erel}$ is the only free parameter of
    Eq.~\eqref{eq:dndtthermalnum}, the estimated value was extracted
    by a single parameter fit, and agrees reasonably well with an
    independent temperature estimate of $3.4\pm 0.3\,\text{K}$
    extrapolated from heating rates measured in the same trap by use
    of the Raman sideband technique
    \cite{epstein07:simpl_ion_heatin_rate_measur}.  }
  \label{fig:realdata}
\end{figure}
The fluorescence predicted by Eq.~\eqref{eq:dndtthermalnum} has been
found to be in good agreement with experimentally observed
fluorescence. We show one experimental data set for comparison in
Fig.~\ref{fig:realdata}; the experiments are more fully described in
\cite{epstein07:simpl_ion_heatin_rate_measur}.
Furthermore, the resulting estimated heating rates have been found to
agree well with results obtained using the Raman sideband technique
\cite{seidelin06:microf_surfac_elect_trap_scalab,epstein07:simpl_ion_heatin_rate_measur}.
This agreement may at first seem surprising, given that the two
methods probe very different energy scales. For the measurements based
on the Raman sideband technique the ion was only allowed to heat for a
few milliseconds thereby gaining a few motional quanta while the
measurement results presented in Fig.~\ref{fig:realdata} are based on
$25\,\text{s}$ heating periods allowing the ion to gain many motional
quanta.  However, the results should agree if, as expected, the
heating rate is constant over these energy scales.

\section{Optimal experimental parameters}
\label{sec:optim-exper-param}

We now examine how the total measurement time required to reach a
given accuracy on the heating rate estimate depends on the choice of
experimental parameters.
As the recoil parameter $r$ will be fixed by choice of atom, we
consider only the choice of optimal values for $\bar{\erel}$, $\drel$,
and laser beam intensity.

It is clear from Fig.~\ref{fig:detuning} that the relevant size of the
signal, in terms of fluorescence photons emitted, for a given initial
motional energy increases with decreasing detuning: the re-cooling is
slower and the change in scattering rate is larger.
For a given experimental setup, the optimal detuning is
decided as a compromise between re-cooling signal and ability to
re-cool atoms that have been highly excited by e.g., collisions.

For a given value of $\bar{\erel}$, the experimental signal, in terms
of the number of photons scattered before steady state is reached,
does not depend on the laser beam intensity.  Since $\bar{\erel}$ is
the average initial energy relative to $E_0$, which is proportional to
the power-broadened linewidth, a lower laser beam intensity will give
a larger signal for a given heating period. This suggests using the
smallest feasible laser intensity, requiring a compromise with respect
to robust cooling and detector dark counts.  From this standpoint we
want to keep the saturation parameter below, but probably close to,
$1$.

For a given detuning and laser intensity, an additional choice of the
length of the heating period in each experimental cycle has to be
made: Should we perform a relatively low number of cycles with long
heating periods or more cycles with shorter heating periods?

\begin{figure}
  \centering
  \includegraphics[width=\linewidth]{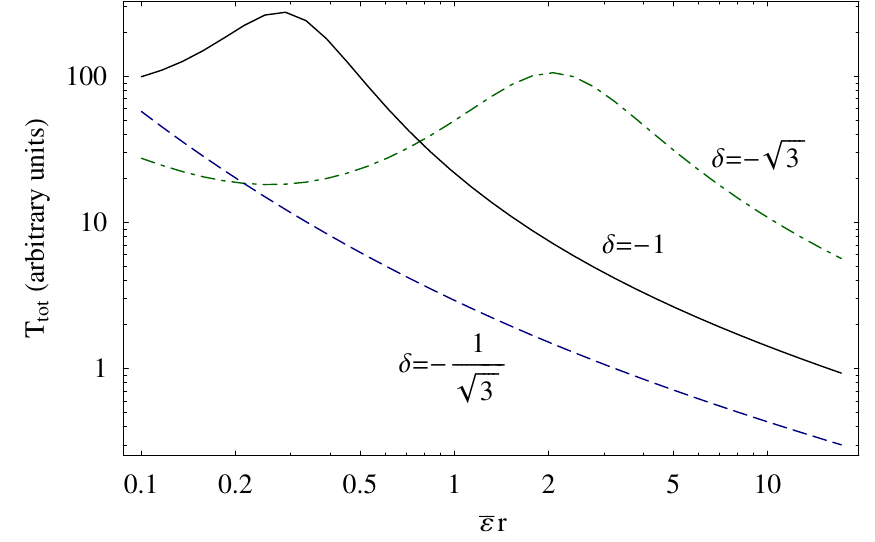}
  \caption{
    Total measurement time required to establish the heating rate with a
    given accuracy, assuming the integration time to be dominated by
    time for re-heating.
    A local maximum is observed for all detunings, but for
    $\drel=-1/\sqrt{3}$ it is located outside the range covered by
    the plot.  
  }
  \label{fig:runtime}
\end{figure}

To answer this question, we estimate the total measurement time,
$\tint$, required to reach a certain relative accuracy on the estimate
of the heating rate.
We assume a constant heating rate and assume that the total time is
dominated by the heating periods, so that $\tint$ is proportional to
the average initial energy, $\bar{E} \propto \bar{\erel}\,\sqrt{1+s}$,
and to the the number of runs.

We will consider a setup where the observed fluorescence is collected
in sequential time-bins that are short compared to the total time
required for the cooling process.
In the limit where the distribution of the integrated number of
counts, $n_i$, in time-bin $i$ is described by a normal distribution
with variance $\sigma_i$, we can estimate the uncertainty on the
maximum-likelihood estimate of $\bar{\erel}$ for a given dataset by
\cite{NR}
\begin{equation}
  \label{eq:uncert}
  1/\sigma(\bar{\erel})^2 = \sum_i \left( \frac{\partial n_i}{\partial
      \bar{\erel}} \right)^2 / \sigma_i^2.
\end{equation}
It follows from Eqs.~\eqref{eq:dedtres}, \eqref{eq:dndtrel}, and
\eqref{eq:dndtthermalnum}, that in the 1-D case the cooling dynamics
can be rewritten in a form independent of $\rrel$ by reparametrizing
in terms of $N\rrel$, $\erel \rrel$, and $\trel \rrel$. We will denote
the reparametrized scattering rate by
\begin{equation}
  \label{eq:dndtavgform}
  \rbar_\drel(\bar{\erel} \rrel,\trel \rrel)=
  \left\langle\frac{\partial (N \rrel)}{\partial (\trel \rrel)}
  \right\rangle_{\bar{\erel} \rrel}.
\end{equation}
Since the relative uncertainty on the heating rate estimate is equal
to $\sigma(\bar{\erel})/\bar{\erel}$ and $\sigma_i=\sqrt{n_i}$,
Eqs.~\eqref{eq:uncert} and \eqref{eq:dndtavgform} allow us to estimate
the time required to obtain a given relative uncertainty on the
heating rate:
\begin{equation}
  \label{eq:runtime}
  \frac{\tint}{\sqrt{1+s}}  \propto 
  \left(
    \bar{\erel} \rrel
    \int_0^\infty 
    \left(
      \frac{\partial \rbar_\drel(\bar{\erel} \rrel,q)}{\partial \bar{\erel} \rrel}
    \right)^2
    \frac{dq}{\rbar_\drel(\bar{\erel} \rrel,q)}
  \right)^{-1}.
\end{equation}
Note that the right hand side depends only on $\delta$ and
$\bar{\erel}\rrel$. Fig.~\ref{fig:runtime} shows $\tint/\sqrt{1+s}$
calculated for different detunings.
The figure confirms that a low detuning is indeed favorable, and also
shows that for a given detuning, $\tint$ decreases with increasing
$\bar{\erel}$. 
This is not surprising, given that the time to cool by a
certain amount of energy increases with atom temperature, as
illustrated by Fig.~\ref{fig:singleion}. 
It is clear from Fig.~\ref{fig:runtime} that the heating period should
be chosen long enough to get a significant signal, i.e.,
$\bar{\erel}>\eofmaxdedt$, but the optimal heating period must be
decided based on other experimental parameters such as trap depth and
background gas collision rate.

\section{Cooling in three dimensions}
\label{sec:effects-spect-modes}

So far, we have considered only cooling in one dimension.  In this
section we will consider the effect of the vibrational modes in other
directions on the cooling process.
Our goal is to gain a qualitative understanding of the effects of
the transverse modes on the cooling dynamics of the $z$ mode, with the
intent of establishing to what extent the simple 1-D model presented
above is a reasonable approximation.

\subsection{3-D cooling of neutral atoms}
\label{sec:3-d-cooling}

For a neutral atom, the confinement transverse to $z$ is not
associated with micromotion, as it is for ions, and the 1-D
weak-binding model extends immediately to three dimensions.
Let $\erel_i$, $i=\{x,y,z\}$, denote the motional energy in mode $i$,
$\dreldi{i}=-\hbar k_i v_i/E_0$ the Doppler shift, and $\dmaxi{i}=2
\sqrt{\erel_i \rrel_i}$ the maximum Doppler shift.
Although all modes are formally identical in the absence of
micromotion, we will discuss the cooling dynamics with a focus on the
$z$ mode.  

\begin{figure}
  \centering
  \includegraphics[width=0.9\linewidth]{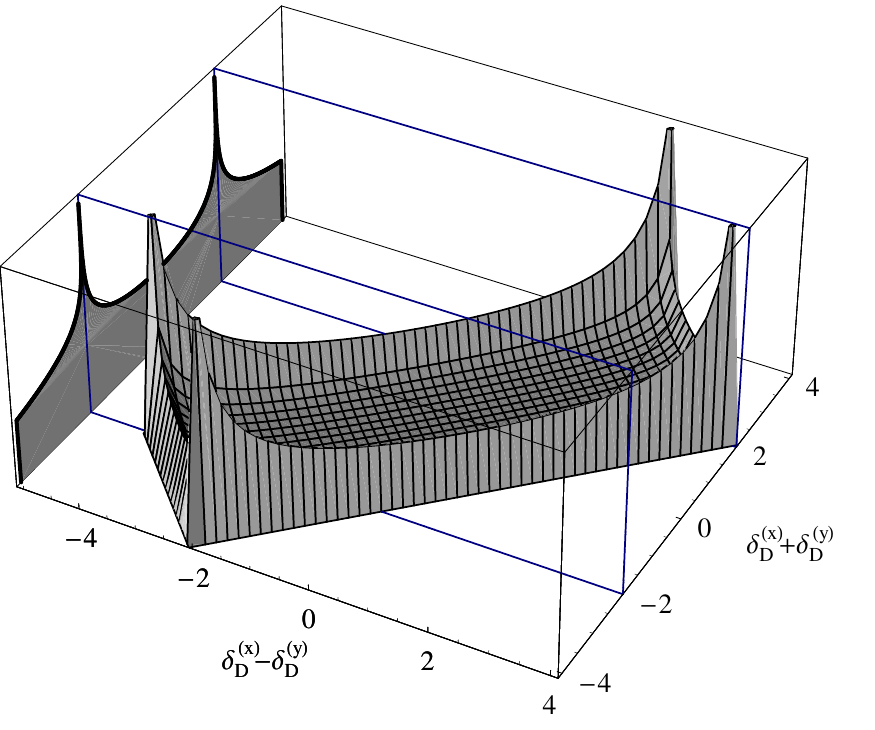}
  \caption{
    The probability density \eqref{eq:ptwoho} for the combined
    Doppler shift $\dreld'=\dreldi{x}+\dreldi{y}$ due to two excited
    modes (curve on left wall) is a marginal distribution of the joint
    probability density
    $\pho(\dmaxi{x},\dreldi{x})\pho(\dmaxi{y},\dreldi{y})$ of
    $(\dreldi{x},\dreldi{y})$ (3-D surface).
    Note that only two of the peaks in the joint probability
    distribution lead to peaks in the marginal distribution.
    Plot is drawn for $\dmaxi{x}=3$ and $\dmaxi{y}=1$, as the
    dash-dotted line in Fig.~\ref{fig:twospectprob}, and
    the joint distribution is truncated to $|\dreldi{i}|<0.95\,
    \dmaxi{i}$ for illustrational purposes.  
  }
  \label{fig:dopplermarginal}
\end{figure}

\begin{figure*}
  \centering
  \belowline{0.45\linewidth}{%
    \belowlabel{(a)}%
    \belowline{\linewidth}{%
      \subfigure{\label{fig:twospectprob}%
        \includegraphics[width=\linewidth]{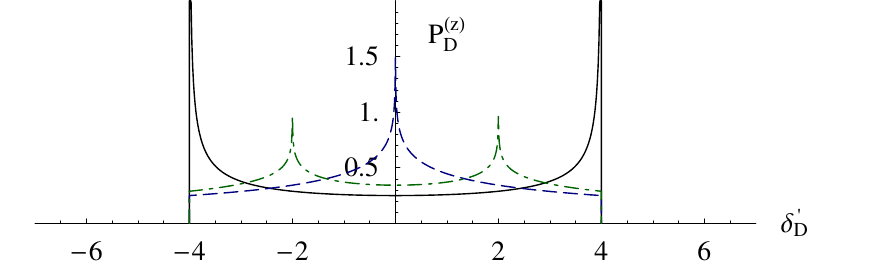}}%
    }
    \belowlabel{(b)}%
    \belowline{\linewidth}{%
      \subfigure{\label{fig:twospectline}%
        \includegraphics[width=\linewidth]{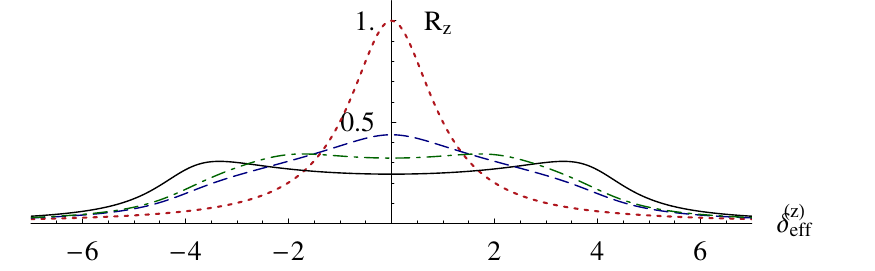}}
    }
  }%
  \belowline{0.5\linewidth}{%
    \belowlabel{(c)}%
    \belowline{\linewidth}{%
      \subfigure{\label{fig:twospectmain}%
        \includegraphics[width=\linewidth]{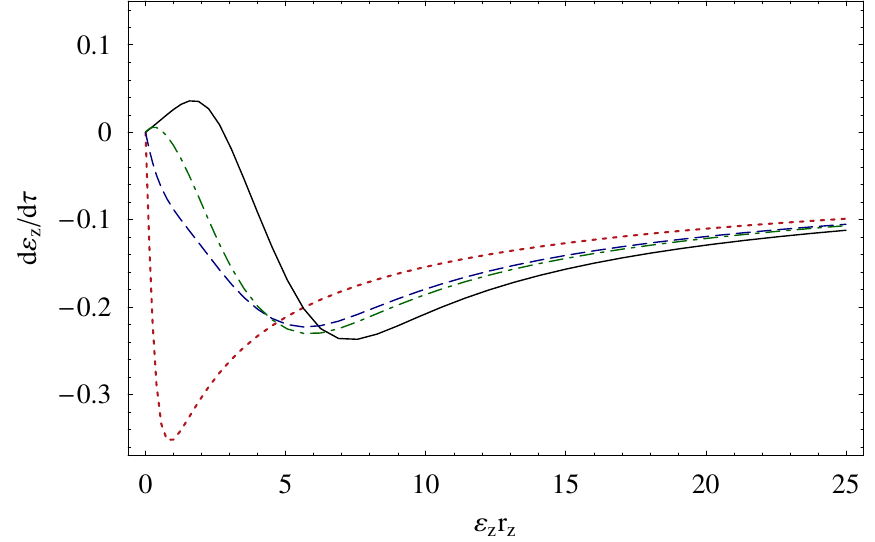}}
    }
  }
  \caption{ \label{fig:Twospect}
    Cooling of the $z$ mode with two excited spectator modes in the
    absence of micromotion.
    Part (a) shows the probability density $\pdz$ of the combined
    Doppler shift due to the $x$ and $y$ modes for
    $(\dmax\topp{x},\dmax\topp{y})$ equal to $(0,4)$ (solid), $(2,2)$
    (dashed), and $(3,1)$ (dash-dotted).
    Part (b) shows the effective line profile $R_z(\deff\topp{z})$
    obtained by convolving $\pdz$ with the Lorentz line profile. Here
    $\deff\topp{z}=\drel+\dreldi{z}$. The spectator mode parameters
    are the same as for (a), and the dotted line show the 
    Lorentz line profile, corresponding to the spectator modes being
    cold, essentially the 1-D cooling case.
    Part (c) shows the $z$ mode cooling rate as a function of
    $\erel_z$ for the effective line profiles of (b).  
  }
\end{figure*}

In experiments it is typically easy to make the frequencies of the
three modes incommensurate, which we will assume here.  In that case,
we can write the rate of change of $\erel_z$ as
\begin{align}
  \label{eq:dedt3d}
  \frac{d\erel_z}{d\trel}
  &=\int \frac{-\dreldi{z}}{1+\left(\drel+\sum_j \dreldi{j}\right)^2}
  \prod_l\pho(\dmaxi{l};\dreldi{l}) d^3\vec{\drel}_D \notag\\
  &=  \int -\dreldi{z}   \pho(\dmaxi{z};\dreldi{z})   R_z(\drel+\dreldi{z})  d\dreldi{z},
\end{align}
where $R_z$ is the effective line profile experienced by the $z$ mode,
obtained by convolving the Lorentz line profile with the distribution
$\pdz$ of the combined Doppler shift $\dreld'\equiv\dreldi{x}+\dreldi{y}$
due to the $x$ and $y$ ``spectator'' modes,
\begin{equation}
  \label{eq:ptwoho}
  \pdz(\dreld')=\int \pho(\dmaxi{x};u) \pho(\dmaxi{y};\dreld'-u) du.
\end{equation}
As illustrated in Figs.~\ref{fig:dopplermarginal} and
\ref{fig:twospectprob}, $\pdz$ is peaked (diverges) at
$\dreld'=\pm|\dmaxi{x}-\dmaxi{y}|$.
If $|\dmaxi{x}-\dmaxi{y}|>2$, the peaks are separated by
more than the width of the Lorentz profile, and $R_z$ will be
double-peaked, as illustrated in Fig.~\ref{fig:twospectline}.
It follows from Eq.~\eqref{eq:dedt3d} that the cooling rate in the
limit of small $\dmaxi{z}$ is proportional to the slope of $R_z$ at
$\drel$, and that the rate of change of $\erel_z$ is positive if the
slope is negative.  If $-|\dmaxi{x}-\dmaxi{y}|<\drel<0$, this will
result in heating of the $z$ mode, at least as long as
$\drel\pm\dmaxi{z}$ are both inside the peaks of $R_z$, i.e., while
$\dmaxi{z}<|\dmaxi{x}-\dmaxi{y}|+\drel$, as illustrated in
Fig.~\ref{fig:twospectmain}
for $(\dmaxi{x},\dmaxi{y})=(0,4)$ and $(1,3)$
\cite{devoe89:role_laser_dampin_trapp_cryst,peik99:sideb_coolin_ions_in_radio}.
The figure also shows that this thermalization or energy equilibration
effect is not present if $|\dmaxi{x}-\dmaxi{y}|\lesssim 1$, as $R_z$
is not double-peaked in this case.
Mathematically, $d\erel_z/dt$, $R_z$, and $\pdz$ are all conveniently
expressed as convolution integrals of functions with known Fourier
transforms.

\begin{figure*}
  \centering
  \belowline{0.48\linewidth}{%
    \belowlabel{(a)}%
    \belowline{\linewidth}{%
      \subfigure{\label{fig:simrates}\includegraphics[width=\linewidth]{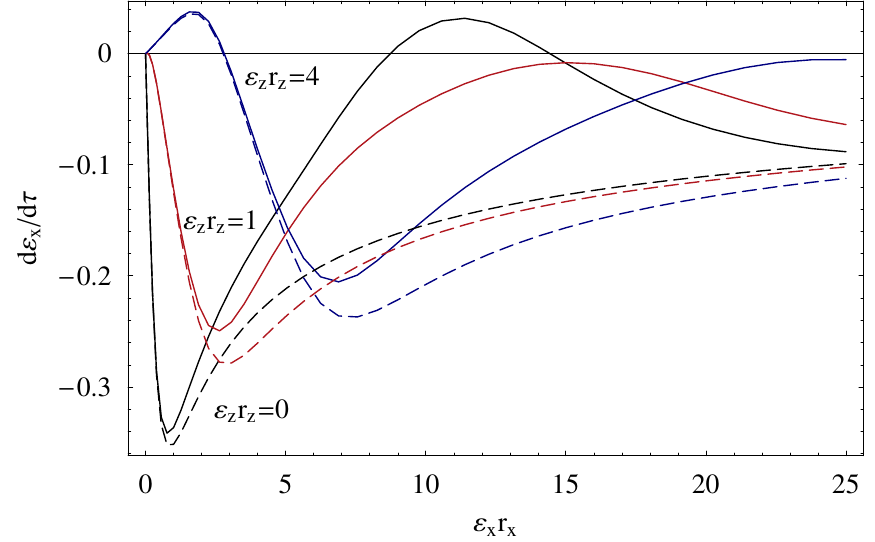}}%
    }
  }\hfill%
  \belowline{0.48\linewidth}{%
    \belowlabel{(b)}%
    \belowline{\linewidth}{%
      \subfigure{\label{fig:simratesax}\includegraphics[width=\linewidth]{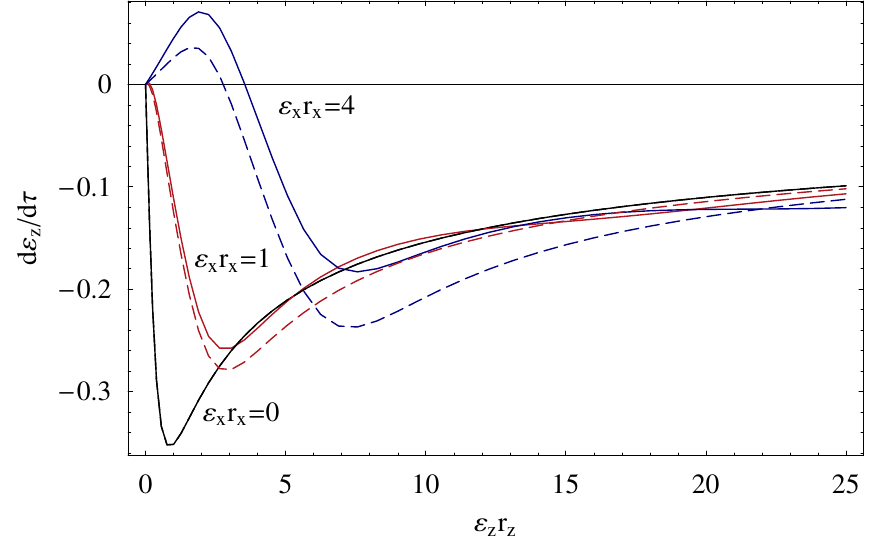}}
    }
  }
  \caption{%
    \label{fig:simratesboth}%
    Predicted cooling rates for (a) a transverse mode in the presence of
    an excited $z$ mode and (b) the $z$ mode in the presence of a
    single excited transverse mode at $\drel=-1$.  
    Dashed lines assume that the trapping potential has no
    associated micromotion, as described by Eq.~\eqref{eq:dedt3d}. In
    this case, the cooling rates are of course identical for the two
    cases.
    The solid lines assume the transverse confinement to be provided
    by an RF quadrupole potential with a frequency $\tilde{\Omega}=6$,
    as described by Eq.~\eqref{eq:dedt3dmm}.
    For both situations, the cooling rate is plotted with the
    spectator mode having a motional energy of $\erel \rrel =
    \{0,1,\text{and }4\}$.
    The plots are based on the weak binding model, and thus assume
    secular frequencies to be small compared to the linewidth.  Note
    that the thermalization/equilibration effect discussed in the text
    is clearly observed for both cases.  
  }
\end{figure*}

The dashed lines in Fig.~\ref{fig:simratesboth} show the cooling rates
predicted by Eq.~\eqref{eq:dedt3d} for the case of only one excited 
spectator mode at different energies. 
When $\dmaxi{z}$ is large compared to $\dmaxi{x}$, we see that the
cooling rate is almost unaffected by the spectator mode. This can be
understood by noting that in the limit where
$\pho(\dmaxi{z};\dreldi{z})$ is uniform over the values of
$\dreldi{z}$ where $R_z(\drel+\dreldi{z})$ is nonzero, the symmetry
of $R_z$ implies that the average energy change per scattering event
is $\drel$, as also discussed in Sec.~\ref{sec:analyzis}.
Since $R_z(\deff)\approx 0$ for $\deff>1+\dmaxi{x}+\dmaxi{y}$, this
implies that the temperature of the spectator modes will not affect
the cooling rate in this limit.
At lower values of $\dmaxi{z}$, we generally see a decrease in the
cooling rate in a gradual approach to the thermalization regime
discussed above.

The consequences of thermalization/equilibration process are complex,
when considering the full 3-D cooling problem. Consider for instance
the case where only one mode is initially hot. According to the
discussion above, this will result in heating of the two remaining
modes, until the fastest heating mode has reached a value of $\dmax$
similar to that of the initially hot mode. After this thermalization,
the modes will be cooled simultaneously at a cooling rate
significantly lower than the cooling rate for a single hot mode.

At this point, it is worth reconsidering the validity of our omission
of recoil heating: The recoil heating rate as given by
Eq.~\eqref{eq:recoil} is seen to have a maximum value of $4 \rrel/3$
at the resonant scattering rate. It is clear from
Fig.~\ref{fig:simratesboth} that for typical values of $\rrel$ on the
order of $10^{-3}$, recoil is insignificant at high energies.

\subsection{3-D cooling of ions including the effects of micromotion}
\label{sec:3d-cooling-ions}

For an ion in a linear Paul trap, if we take the $z$ direction to be the
axis, confinement in the transverse $x$ and $y$ directions is provided
by the ponderomotive potential of an RF quadrupole field.
The full 3-D cooling problem including micromotion on the transverse modes
is very complex even in the Lamb-Dicke regime
\cite{devoe89:role_laser_dampin_trapp_cryst,walther93:phase_trans_stored_laser_cooled,cirac94:laser_coolin_trapp_ions}.
At low saturation, the effects of the micromotion caused by an RF
field of frequency $\Omega$ can be modeled by including micromotion
sidebands in the line profile
\cite{devoe89:role_laser_dampin_trapp_cryst}:
Micromotion with peak amplitude $\vec{\amm}(\xavg(t))$, where
$\xavg$ is the ion position averaged over one period of the RF
field, can be described by the line profile,
\begin{equation}
  \label{eq:rhoeemm}
  \reerel(\deff,\beta)= \sum_{n=-\infty}^\infty J_n^2(\beta) \, \frac{1}{1+(\deff-n \tilde{\Omega})^2},
\end{equation}
where $\deff=\drel+\sum_j \dreldi{j}$ is the effective detuning, 
$\beta=\left\vert\vec{\amm}(\xavg)\cdot\vec{k}\right\vert$ is the
micromotion modulation index, $\tilde{\Omega}=\hbar \Omega/E_0$ is
the scaled RF frequency, and $J_n$ is the $n$-th Bessel function.

In contrast to the situation in
Ref.~\cite{devoe89:role_laser_dampin_trapp_cryst}, we are considering
a case where $\beta$ changes during the secular motion.
Since $\beta$ and $\deff$ depend on $\xavg$ and $\dot{\xavg}$,
respectively, we parametrize the secular motion by the instantaneous
phases, $\phi_i$, where $\xiavg=\xiavg\topp{0} \cos(\phi_i(t))$, and
where $\xiavg\topp{0}$ is slowly varying and $\dot{\phi}_i\approx
\omega_i$.
Choosing the $x$ and $y$ axes so that the RF field is proportional to
$(\bar{x} \vhat{x}-\bar{y} \vhat{y}) \cos(\Omega t)$, we find that
$\dreldi{i}=\dmaxi{i} \sin(\phi_i)$.
In the limit where the transverse confinement is modified only weakly by
static potentials, so that $\omega_x \approx \omega_y$, we find in the
pseudopotential approximation that 
$\beta=\sqrt{2} \left\vert \dmaxi{x}
  \cos(\phi_x)-\dmaxi{y}\cos(\phi_y) \right\vert/\tilde{\Omega}$, which
we note to be independent of the secular frequencies.
%
In this case we have
\begin{equation}
  \label{eq:dedt3dmm}
  \frac{d\erel_i}{d\trel}=-\int \drel_i\,
  \reerel(\deff(\vec{\phi}),\beta(\phi_x,\phi_y))
  \frac{d^3\vec{\phi}}{(2\pi)^3},
\end{equation}
where the integral is over $[0,2\pi]$ in all dimensions.
Note that since the modulation index depends only on the transverse
components of the motion, the effect of excited transverse modes on the
cooling of the $z$ mode can still be described in terms of an
effective line profile, as in Eq.~\eqref{eq:dedt3d}. 

For the cooling of the transverse modes, the effects of micromotion on
the cooling rates is pronounced, as illustrated by
Fig.~\ref{fig:simrates}.
%
A very clear qualitative difference from the cooling rate in the
micromotion-free case is that at sufficiently high RF frequencies
($\tilde{\Omega}>4.4$ for $\drel=-1$), stable points for the transverse
mode energies develop even when the remaining modes are cold. 
This effect has been discussed in
Ref.~\cite{peik99:sideb_coolin_ions_in_radio} and is attributed to the
heating peak of the Doppler distribution becoming resonant with a
micromotion sideband, as described by Eq.~\eqref{eq:rhoeemm}.
This might be related to the bistable behavior reported in some single
ion experiments
\cite{sauter88:kinet_singl_trapp_ion,sauter88:photo_dynam_singl_ions_in,walther93:phase_trans_stored_laser_cooled}.
The stability breaks down when thermalization is taken into
consideration. Consider for instance the stable point indicated by
Fig.~\ref{fig:simrates} to exist for
$\vec{\erel}\approx(15/\rrel_x,0,0)$. Here, it is clear from the
figure that when the $z$ mode has heated to $\erel_z>1/\rrel_z$,
cooling of the $x$ mode will commence.

When $\dmax/ \tilde{\Omega} \lesssim \sqrt{2}$ for the transverse
modes, we find that only the $J_0$ term of Eq.~\eqref{eq:rhoeemm}
contributes significantly, and the argument of
Sec.~\ref{sec:3-d-cooling} that the cooling rate for the $z$ mode is
not affected by excited transverse modes when
$\dmaxi{z}>1+\dmaxi{x}+\dmaxi{y}$ also applies here, as illustrated by
Fig.~\ref{fig:simratesax}.

It is clear from the results above that we cannot ignore the transverse
modes if their associated maximal Doppler shifts are comparable to
that of the $z$ mode.
If, however, we assume the transverse modes are cold enough to avoid
the heating effects described in Figs.~\ref{fig:Twospect} and
\ref{fig:simratesboth}, we have seen above that the primary effect of
the transverse modes will be to slow down the cooling of the $z$ mode.
This would result in the 1-D model overestimating the mean initial
energy of the $z$ mode.
However, for many experiments that use linear RF traps, it is
reasonable to assume that the transverse modes are heated
significantly less than the $z$ mode. This is because most
investigations of the anomalous heating in ion traps have found the
results to be consistent with heating rates having a frequency
dependence of $\omega^{-n}$ with $n>1$
\cite{turchette00:heatin_trapp_ions_from_quant,deslauriers06:scalin_suppr_anomal_quant_decoh,epstein07:simpl_ion_heatin_rate_measur}.
Since the transverse mode frequencies are often an order of magnitude
larger than $\omega_z$, this would indeed lead to the transverse modes
being significantly colder than the $z$ mode.
Also, since the energy in the transverse modes only affects the
cooling of the $z$ mode through the resulting Doppler shift, the
effect of the transverse modes could be further reduced by aligning
$\vec{k}$ to have a smaller projection on the transverse modes. This
would however reduce the efficiency of cooling of the transverse modes
\cite{itano82:laser_coolin_ions_stored_in}.

Finally, another effect with respect to micromotion is that the presence of
uncontrolled static stray fields can result in the ion experiencing
micromotion even at the ion equilibrium position. At temperatures
where $\dmax\ll \tilde{\Omega}$, the first order effect according to
Eq.~\eqref{eq:rhoeemm} of this will be a reduction of the central
spectral component by a factor of $J_0(\beta)^2$; see for example
Ref.~\cite{berkeland98:minim_microm_a_paul_trap}.
We note that this effect can be compensated by using an
effective saturation parameter based on the steady-state fluorescence
observed in the trap.

\begin{figure}
  \centering
  \includegraphics[width=\linewidth]{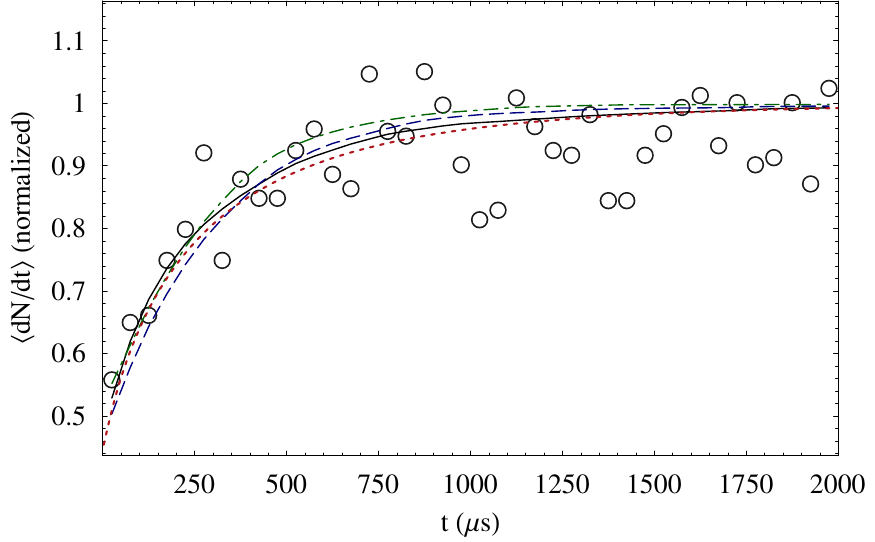}
  \caption{ 
    Averaged fluorescence vs.\ time as predicted by Monte Carlo simulations
    for best fits to the data presented in Fig.~\ref{fig:realdata} for
    different models of the frequency dependence of
    the anomalous heating rate:
    The solid line assumes transverse modes to be unaffected by the heating;
    this should be very well approximated by the 
    best fit to the 1-D model (dotted), as also
    plotted in Fig.~\ref{fig:realdata}.
    For the dashed and dash-dotted lines, we assume heating to be
    proportional to $\omega^{-1.4}$ and independent of $\omega$,
    respectively. 
    The fitted value of the $z$ mode temperature for the three cases
    is $3.9\,\text{K}$, $3.7\,\text{K}$, and $1.4\,\text{K}$  }
  \label{fig:simulation}
\end{figure}

\subsection{Departures from the weak-binding, low-saturation limit}
\label{sec:montecarlo}

In most experimental situations, we will not strictly fulfill the
requirements of low saturation or weak binding. In particular, for
the trap referenced in Fig.~\ref{fig:realdata} the secular frequencies
of the transverse modes are approximately equal to half the $41.4\,\text{MHz}$
linewidth of the Doppler cooling transition, making the weak binding
assumption only approximate. Also, the illustrated data were obtained at
a saturation parameter of $0.9$, outside the validity region of the
line-profile model that accounts for RF micromotion \eqref{eq:dedt3dmm}.
To validate our claim that the fluorescence signal predicted by the
1-D model is a good approximation if the heating rate is assumed to be
a strongly decreasing function of $\omega$, we performed a
numerical Monte Carlo simulation of the fluorescence, based on
integrating the optical Bloch equations through a large number of
cooling trajectories.
For each trajectory, we propagate the density matrix $\rho$ of the
ion's internal state according to the master equation
\begin{equation}
  \label{eq:master}
  \frac{d \rho}{d t}=
  \frac{i}{\hbar}\left[\rho,H\topp{c}(\vec{x},t)\right]
    + 2L \rho L^\dag -\left\{L^\dag L, \rho\right\},
\end{equation}
where $L\equiv \kg\be\,\sqrt{\Gamma/2}$ is the Lindblad operator for
excited state decay and $\vec{x}(t)=\xavg(t)+\vec{a}(\xavg(t))\cos(
\Omega t)$ for the $\xavg$ and $\vec{a}$ introduced above.
Coupling to the motional state is modeled by the average light force,
$m \ddot{\xavg}=\hbar \vec{k}\, \Gamma\, \ree(t)$.
This model assumes neither the atoms to be weakly bound nor the
cooling beam intensity to be low but does neglect recoil heating.

Figure \ref{fig:simulation} shows the result of fitting simulations
with different assumptions for the frequency dependence of the heating
to the dataset presented in Fig.~\ref{fig:realdata}.
We find that if we assume the transverse modes are not heated, we
obtain a temperature estimate of $3.9\,\text{K}$, in agreement with
the result of fitting the 1-D model to the data, as illustrated by
Fig.~\ref{fig:realdata}.
%
The $z$ mode temperature of $3.7\,\text{K}$ estimated from the
$\omega^{-1.4}$ model is close to, and slightly smaller than, this
value, and agrees with the temperature estimate of $3.4\pm
0.3\,\text{K}$ based on extrapolating heating rates measured with the
Raman sideband technique for the same trap configuration. 
This
particular form of the frequency dependence of the heating rate was
observed for the same trap when the Raman sideband technique
\cite{epstein07:simpl_ion_heatin_rate_measur} was used, and similar
frequency dependencies have been observed in other geometries
\cite{turchette00:heatin_trapp_ions_from_quant,deslauriers06:scalin_suppr_anomal_quant_decoh}.
If we instead assume an $\omega^{-1}$ dependence of the heating, the results
only change slightly.

Our main conclusions from the simulation results are that the primary
effect of the presence of weakly heated spectator modes will be to
slow down cooling due to thermalization. If $\omega^{-1.4}$ heating of
the transverse modes is assumed, the 1-D model will somewhat
overestimate the motional temperature of the axial mode.

\section{Modified experimental protocols}
\label{sec:modif-exper-prot}

We consider two modifications to the experimental protocol to reduce
the total measurement time.  Both are motivated by the fact that the
size of the signal from a given amount of heating increases with
increased initial energy.

\begin{figure}
  \centering
  \includegraphics[width=\linewidth]{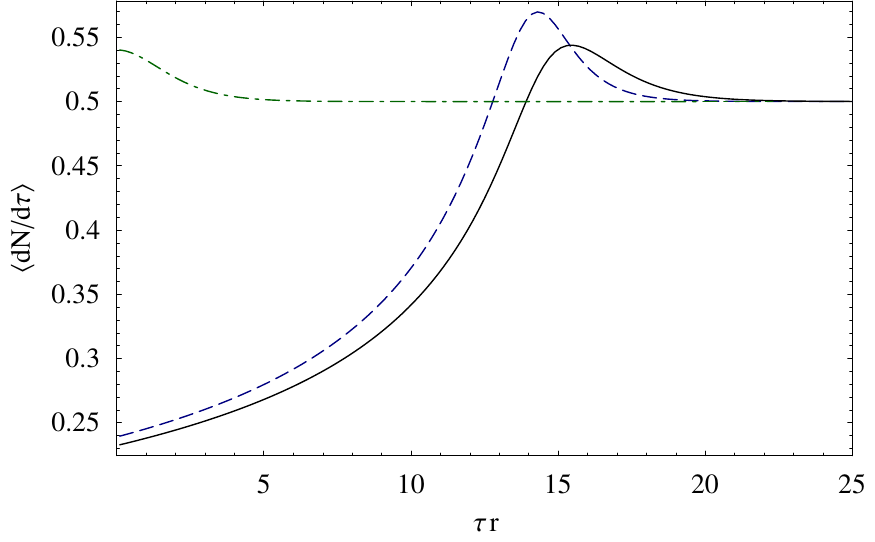}
  \caption{
    Improving the sensitivity of the temperature measurement by
    deliberate excitation. 
    The dashed line shows the average
    scattering rate \eqref{eq:dndtthermalnum} as a function of time
    for an atom that was
    deliberately excited to a motional energy of $\erel_0=5 \eofmaxdedt$,
    so that $P_0(\erel)=\delta_\text{Dirac}(\erel-\erel_0)$.
    The solid line shows the scattering rate as a function of time for
    an atom which has first been deliberately excited to a motional
    energy of $\erel_0=5 \eofmaxdedt$ and then allowed to heat for a duration which added
    an average thermal energy of $\bar{\erel}=0.25/r$.
    For comparison, the dash-dotted line
    shows the signal for an atom experiencing the same heating
    period without any initial excitation, i.e.~with $\bar{\erel}=0.25/\rrel$.  
    In all cases, $\drel=-1$.  }
  \label{fig:offsete}
\end{figure}

One approach would be to coherently add a known amount of energy to
the $z$ mode at the start of the heating period. If the added energy
is enough to bring the atom into the slow-cooling regime, this will
increase the signal change due to a given amount of additional
heating, as illustrated in Fig.~\ref{fig:offsete}.

\begin{figure}
  \centering
  \includegraphics[width=\linewidth]{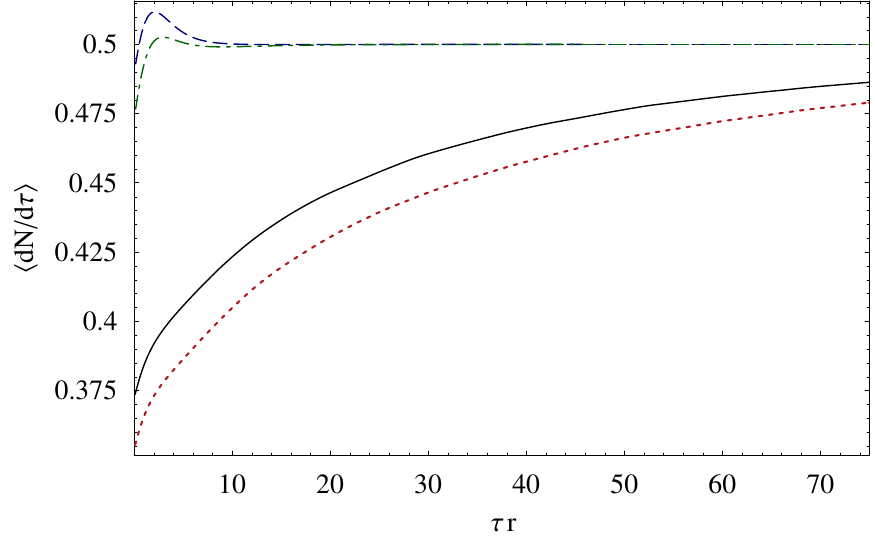}
  \caption{Improving the sensitivity of the temperature measurement by
    parametric amplification.
    The dashed and dash-dotted lines show the average scattering rate
    vs.~time during the re-cooling of an atom which
    has been allowed to heat to $\bar{\erel}=0.9 \eofmaxdedt$ and $1.1
    \eofmaxdedt$ respectively at $\drel=-1$.
    The solid and dotted lines show the average scattering rate for
    the same conditions, except that the motional amplitude has
    been parametrically amplified by a factor of $3$ after the heating
    period.
  }
  \label{fig:paramamp}
\end{figure}

Alternatively, parametric amplification
\cite{dehmelt68:bolom_techn_for_rf_spect,caves82:quant_limit_noise_in_linear,heinzen90:quant_limit_coolin_and_detec}
could be employed after the heating cycle to modify the thermal
distribution. Parametric amplification can be implemented by
modulating the $z$ trap potential at $2 \omega_z$, and leads to
amplification of one quadrature of the motion while damping the other
quadrature.
For a low value of $\bar{\erel}$, parametric amplification would
increase the fraction of experiments in which the atom is in the
slow-cooling regime at the beginning of the cooling process, thus
increasing the signal for a given heating period, as illustrated in
Fig.~\ref{fig:paramamp}.

\section{Conclusion}
\label{sec:conclusion}

In conclusion, we have shown that the motional energy of a trapped
atom or ion can be estimated from the temporal changes in fluorescence
observed when Doppler cooling is applied. Specifically, the initial
energy can be estimated by fitting Eq.~\eqref{eq:dndtthermalnum},
where the mean initial motional energy is the only free variable, to
the observed fluorescence.
Our analysis assumes the oscillation frequency of the atoms is much
smaller than the linewidth of the optical transition used for Doppler
cooling and the motional energy at the start of the cooling is
thermal.

Compared to Raman sideband transition methods for heating rate
measurements, this method is simpler to implement experimentally but
requires longer measurement duration for traps with low heating
rates. On the other hand, for high heating rates, where sideband
cooling is inefficient, this may be the method of choice.
We have shown that in the typical situation, where the time for
heating dominates, the total measurement time decreases with
decreasing laser intensity, decreasing laser detuning, and increased
heating period duration. We have compared the trade-off between these
parameters (Fig.~\ref{fig:runtime} and Eq.~\eqref{eq:runtime}).
Finally, we show (Sec.~\ref{sec:optim-exper-param}) that the total
measurement time can be reduced by adding additional energy to more
quickly bring the ion into the low fluorescence regime.

By comparison with various models of 3-dimensional Doppler cooling, we
have established that under typical experimental conditions the
effects of the high-frequency modes are small, and that they will lead
to temperature estimates that are somewhat higher than the actual
temperature of the low-frequency mode.

Work supported by DTO and NIST.
J.H.W. acknowledges support from The Danish Research Agency. 
R.J.E. acknowledges National Research Council Research Associateship Awards.
S.S. acknowledges support from the Carlsberg Foundation. 
J.P.H. acknowledges support from a Lindemann Fellowship.
We thank J.~J.~Bollinger and C.~Ospelkaus for comments on the manuscript.
This manuscript is a publication of NIST and is not subject to
U.S. copyright.

\appendix
\section{Integrals}
\label{sec:integrals}
The integrals appearing in Eqs.~\eqref{eq:dedtrel},
\eqref{eq:dndtrel}, and \eqref{eq:dedt3d} are all convolution integrals of
elements with analytical Fourier transforms and can thus be easily
evaluated in Fourier space.
For the 1-D integrals, the inverse Fourier transform can also be
performed analytically. Here we present a more direct approach to
evaluating the 1-D integrals.

For $a,b \in \mathbb{R}$ we define $Z(a,b)$ as
\begin{equation}\label{eq:z-equiv-int_02}
  Z \equiv
  \int_0^{2 \pi} \frac{1}{\sin(\phi)-z} \frac{d \phi}{2 \pi} = -\frac{1}{z}\,
  \sqrt{\frac{z^2}{z^2-1}},
\end{equation}
where $z=(a+i)/b$. Noting that 
\begin{equation}
  \frac{1}{x-z}=b \frac{1}{1+(a-b x)^2} \left((b x-a)+i \right),
\end{equation}
we find that, according to Eq.~\eqref{eq:z-equiv-int_02},
\begin{align*}
  \int^{2 \pi}_0 \frac{b \sin(\phi)}{1+(a-b \sin(\phi))^2} \frac{d
  \phi}{2 \pi} &= \frac{1}{b}\,\left(\re(Z)+a \im(Z)\right)\\
  \int^{2 \pi}_0 \frac{1}{1+(a-b \sin(\phi))^2} \frac{d
  \phi}{2 \pi} &= \frac{1}{b}\,\im(Z).
\end{align*}
Taking the branch cut discontinuity for $\sqrt{\cdot}$ to be along the negative real
axis, we have for $b>0$ that $\sqrt{(-i z)^2}=-i z$, so that
\begin{equation}
  \label{eq:zthiscase}
  Z(a,b)= \frac{i b}{\sqrt{b^2-(a+i)^2}},\qquad b>0.
\end{equation}

\section{Numerical calculation of the averaged scattering rate}
\label{sec:numer-eval-therm}

In this section we present an efficient numerical method for
evaluating the averaged scattering rate given by
Eq.~\eqref{eq:dndtthermalnum}.

Introducing $\erel_n = \eprop(\erel_0,n \Delta \trel)$, for
$n=0,1,\ldots$, we note that $\eprop(\erel_m,n \Delta \trel) =
\erel_{m+n}$. The values of $\erel_n$ are the energies along a single
cooling trajectory.
If the scattering rate can be considered constant on time scales of
$\Delta \trel$,
\begin{equation}
  \label{eq:rdef}
  \frac{dN}{d\trel}(\eprop(\erel_0,\trel))\approx R_n, 
  \qquad \trel \in [(n-1) \Delta \trel,n \Delta \trel],
\end{equation}
we find that the thermally averaged scattering rate, as given by
Eq.~\eqref{eq:dndtthermalnum}, averaged over the same intervals can be
approximated by
\begin{equation}
  \label{eq:rbarconv}
  \bar{R}_n\approx \sum_{m=0}^\infty 
  R_{m+n}\ 
  \int_{\erel_{m}}^{\erel_{m+1}} P_0(\erel') d\erel'. 
\end{equation}
Since the values of the $R_n$ are independent of
$P_0$, $\bar{R}_n$ is easily calculated for different $P_0$ by list
convolution.

For the Maxwell-Boltzmann distribution, a numerically stable form of
the weight factors appearing in \eqref{eq:rbarconv} is
\begin{equation*}
  \int_{\erel-\Delta \erel/2}^{\erel+\Delta\erel/2}
e^{-\erel'/\bar{\erel}} \frac{d\erel'}{\bar{\erel}} =
2 e^{-\erel/\bar{\erel}} \sinh\left(\frac{\Delta \erel}{2 \bar{\erel}}\right).
\end{equation*}


\end{document}